# The symmetries of the Fokker - Planck equation in two dimensions

*Igor A. Tanski*
*tanski@protek.ru*

ZAO CV Protek

*ABSTRACT*

We calculate all point symmetries of the Fokker - Planck equation in two-dimensional Euclidean space. General expression of symmetry group action on arbitrary solution of Fokker - Planck equation is presented.

## 1. The symmetries of the Fokker - Planck equation in two dimensions

The object of our considerations is a special case of Fokker - Planck equation, which describes evolution of 2D continuum of non-interacting particles imbedded in a dense medium without outer forces. The interaction between particles and medium causes combined diffusion in physical space and velocities space. The only force, which acts on particles, is damping force proportional to velocity.

The 3D variant of this equation was investigated in our work [1]. In this work fundamental solution of 3D equation was obtained by means of Fourier transform.

The 1D variant of this equation was investigated in our work [2]. All point symmetries of the Fokker - Planck equation in one-dimensional Euclidean space were calculated.

In present work we continue this investigation for more complex 2D equation.

The Fokker - Planck equation in two dimensions is

$$\frac{\partial n}{\partial t} + u \frac{\partial n}{\partial x} + v \frac{\partial n}{\partial y} - au \frac{\partial n}{\partial u} - av \frac{\partial n}{\partial v} - 2an - k\left(\frac{\partial^2 n}{\partial u^2} + \frac{\partial^2 n}{\partial v^2}\right) = 0; \qquad (1)$$

where

$n = n(t, x, y, u, v)$ - density;

$t$ - time variable;

$x, y$ - space coordinates;

$u, v$ - velocity;

$a$ - coefficient of damping;

$k$ - coefficient of diffusion.

The list of symmetries of the Fokker - Planck equation in one dimension follows. The calculations of symmetries are rather awkward. They are carried out to APPENDIX 1.

Instead of classic "$\xi - \phi$" notation we use another ("$\delta$") notation. This notation was presented in our work [2].

Addition of arbitrary solution

$$\mathbf{v}_1 = A \frac{\partial}{\partial n}; \qquad (2)$$

where *A* is arbitrary solution of the (1) equation.

Scaling of density

$$\mathbf{v}_2 = n \frac{\partial}{\partial n} ; \tag{3}$$

The reason of symmetries (2-3) existence is linearity of PDE (1).

Time shift

$$\mathbf{v}_3 = \frac{\partial}{\partial t} ; \tag{4}$$

Space translations

$$\mathbf{v}_4 = \frac{\partial}{\partial x} ; \quad \mathbf{v}_5 = \frac{\partial}{\partial y} . \tag{5}$$

Space rotation

$$\mathbf{v}_6 = v \frac{\partial}{\partial u} - u \frac{\partial}{\partial v} + y \frac{\partial}{\partial x} - x \frac{\partial}{\partial y} . \tag{6}$$

Transformations (5) and (6) build two-dimensional Euclidean movements group.

Extended Galilean transformations, which besides time and space coordinates affect the density

$$\mathbf{v}_7 = \frac{\partial}{\partial u} + t \frac{\partial}{\partial x} - \frac{an}{2k}(ax+u)\frac{\partial}{\partial n} ; \quad \mathbf{v}_8 = \frac{\partial}{\partial v} + t \frac{\partial}{\partial x} - \frac{an}{2k}(ax+v)\frac{\partial}{\partial n} . \tag{7}$$

Negative exponent transformations - they affect time and space coordinates, contain time-dependent common multiplier (negative exponent). They do not affect density.

$$\mathbf{v}_9 = e^{-at}\left(-a\frac{\partial}{\partial u} + \frac{\partial}{\partial x}\right); \quad \mathbf{v}_{10} = e^{-at}\left(-a\frac{\partial}{\partial v} + \frac{\partial}{\partial y}\right). \tag{8}$$

Positive exponent transformations - they affect time, space and density, contain time-dependent common multiplier (positive exponent).

$$\mathbf{v}_{11} = e^{at}\left(a\frac{\partial}{\partial u} + \frac{\partial}{\partial x} - \frac{a^2}{k}nu\frac{\partial}{\partial n}\right); \quad \mathbf{v}_{12} = e^{at}\left(a\frac{\partial}{\partial v} + \frac{\partial}{\partial y} - \frac{a^2}{k}nv\frac{\partial}{\partial n}\right). \tag{9}$$

One-parameter groups, generated by vector fields $\mathbf{v}_1 - \mathbf{v}_{12}$, are enumerated in the following list. The list contains images of the point $(n, t, x, y, u, v)$ by transformation $exp(\varepsilon \mathbf{v}_i)$

$G_1$:  $(n + \varepsilon A, t, x, y, u, v)$;

$G_2$:  $(e^\varepsilon n, t, x, y, u, v)$;

$G_3$:  $(n, t + \varepsilon, x, y, u, v)$;

$G_4$:  $(n, t, x + \varepsilon, y, u, v)$;

$G_5$:  $(n, t, x, y + \varepsilon, u, v)$;

$G_6$:  $(n, t, \cos(\varepsilon)x + \sin(\varepsilon)y, -\sin(\varepsilon)x + \cos(\varepsilon)y, \cos(\varepsilon)u + \sin(\varepsilon)v, -\sin(\varepsilon)u + \cos(\varepsilon)v)$;   (10)

$G_7$:  $\left(\exp\left[-\frac{a}{2k}\left(\varepsilon(ax+u) + \frac{1}{2}\varepsilon^2(at+1)\right)\right]n, t, x + \varepsilon t, y, u + \varepsilon, v\right)$;

$G_8$: $\quad (\exp\left[-\dfrac{a}{2k}\left(\varepsilon(ay+v)+\dfrac{1}{2}\varepsilon^2(at+1)\right)\right] n, t, x, y+\varepsilon t, u, v+\varepsilon);$

$G_9$: $\quad (n, t, x+\varepsilon e^{-at}, y, u-\varepsilon a\, e^{-at}, v);$

$G_{10}$: $\quad (n, t, x, y+\varepsilon e^{-at}, u, v-\varepsilon a\, e^{-at});$

$G_{11}$: $\quad (n \exp\left[-\dfrac{a^2}{k} e^{at}\left(\varepsilon u+\dfrac{1}{2}\varepsilon^2 a\, e^{at}\right)\right], t, x+\varepsilon e^{at}, y, u+\varepsilon a\, e^{at}, v).$

$G_{12}$: $\quad (n \exp\left[-\dfrac{a^2}{k} e^{at}\left(\varepsilon v+\dfrac{1}{2}\varepsilon^2 a\, e^{at}\right)\right], t, x, y+\varepsilon e^{at}, u, v+\varepsilon a\, e^{at}).$

For relatively nontrivial integration of $G_7$, $G_8$, $G_{11}$ and $G_{12}$ we refer to [2].

The fact, that $G_i$ are symmetries of PDE (1) means, that if $f(t, x, y, u, v)$ is arbitrary solution of (1), the functions

$u^{(1)}$: $\quad f(t, x, y, u, v) + \varepsilon A(t, x, y, u, v);$

$u^{(2)}$: $\quad e^{\varepsilon} f(t, x, y, u, v);$

$u^{(3)}$: $\quad f(t-\varepsilon, x, y, u, v);$

$u^{(4)}$: $\quad f(t, x-\varepsilon, y, u, v);$

$u^{(5)}$: $\quad f(t, x, y-\varepsilon, u, v);$

$u^{(6)}$: $\quad f(t, \cos(\varepsilon)x - \sin(\varepsilon)y, \sin(\varepsilon)x + \cos(\varepsilon)y, \cos(\varepsilon)u - \sin(\varepsilon)v, \sin(\varepsilon)u + \cos(\varepsilon)v);$

$u^{(7)}$: $\quad \exp\left[-\dfrac{a}{2k}\left(\varepsilon(ax+u) - \dfrac{1}{2}\varepsilon^2(at+1)\right)\right] f(t, x-\varepsilon t, y, u-\varepsilon, v);$ (11)

$u^{(8)}$: $\quad \exp\left[-\dfrac{a}{2k}\left(\varepsilon(ay+v) - \dfrac{1}{2}\varepsilon^2(at+1)\right)\right] f(t, x, y-\varepsilon t, u, v-\varepsilon);$

$u^{(9)}$: $\quad f(t, x-\varepsilon e^{-at}, y, u+\varepsilon a\, e^{-at}, v);$

$u^{(10)}$: $\quad f(t, x, y-\varepsilon e^{-at}, u, v+\varepsilon a\, e^{-at});$

$u^{(11)}$: $\quad \exp\left[-\dfrac{a^2}{k} e^{at}\left(\varepsilon u - \dfrac{1}{2}\varepsilon^2 a\, e^{at}\right)\right] f(t, x-\varepsilon e^{at}, y, u-\varepsilon a e^{at}, v).$

$u^{(12)}$: $\quad \exp\left[-\dfrac{a^2}{k} e^{at}\left(\varepsilon v - \dfrac{1}{2}\varepsilon^2 a\, e^{at}\right)\right] f(t, x, y-\varepsilon e^{at}, u, v-\varepsilon a e^{at}).$

where $\varepsilon$ - arbitrary real number, also are solutions of (1). Here $A$ is another arbitrary solution of (1).

We systematically replaced "old coordinates" by their expressions through "new coordinates". Note, that due to these replacements terms with $\varepsilon^2$ in $u^{(7)}$, $u^{(8)}$, $u^{(11)}$ and $u^{(12)}$ change their signs.

We have trivial solution $n = e^{2at}$ at our disposal. If we act on this solution by transformations (10), we obtain 4 new solutions:

$$n = \exp\left[2at - \frac{a}{2k}\left(\varepsilon(ax+u) - \frac{1}{2}\varepsilon^2(at+1)\right)\right]; \tag{12}$$

$$n = \exp\left[2at - \frac{a}{2k}\left(\varepsilon(ay+v) - \frac{1}{2}\varepsilon^2(at+1)\right)\right]. \tag{13}$$

$$n = \exp\left[2at - \frac{a^2}{k}e^{at}\left(\varepsilon u - \frac{1}{2}\varepsilon^2 a\, e^{at}\right)\right]; \tag{14}$$

$$n = \exp\left[2at - \frac{a^2}{k}e^{at}\left(\varepsilon v - \frac{1}{2}\varepsilon^2 a\, e^{at}\right)\right]. \tag{15}$$

General expression is

$$U = e^{\varepsilon_2}\exp\left[-\frac{a}{2k}\left(\varepsilon_7(a\bar{x}+\bar{u}) - \frac{1}{2}\varepsilon_7^2(at+1)\right)\right]\exp\left[-\frac{a}{2k}\left(\varepsilon_8(a\bar{y}+\bar{v}) - \frac{1}{2}\varepsilon_8^2(at+1)\right)\right]\times \tag{16}$$

$$\times \exp\left[-\frac{a^2}{k}e^{at}\left(\varepsilon_{11}(\bar{u}-\varepsilon_7) - \frac{1}{2}\varepsilon_{11}^2 a\, e^{at}\right)\right]\exp\left[-\frac{a^2}{k}e^{at}\left(\varepsilon_{12}(\bar{v}-\varepsilon_8) - \frac{1}{2}\varepsilon_{12}^2 a\, e^{at}\right)\right]\times$$

$$\times f(t-\varepsilon_3, \bar{x}-\varepsilon_4-\varepsilon_7 t-\varepsilon_9 e^{-at}-\varepsilon_{11}e^{at}, \bar{y}-\varepsilon_5-\varepsilon_8 t-\varepsilon_{10}e^{-at}-\varepsilon_{12}e^{at}, \bar{u}-\varepsilon_7+\varepsilon_9 a\, e^{-at}-\varepsilon_{11}a\, e^{at}, \bar{v}-\varepsilon_8+\varepsilon_{10}a\, e^{-at}-\varepsilon_{12}a\, e^{at}) +$$

$$+\varepsilon_1 A(t,x,y,u,v);$$

where

$$\bar{x} = \cos(\varepsilon_6)x - \sin(\varepsilon_6)y; \tag{17}$$

$$\bar{y} = \sin(\varepsilon_6)x + \cos(\varepsilon_6)y; \tag{18}$$

$$\bar{u} = \cos(\varepsilon_6)u - \sin(\varepsilon_6)v; \tag{19}$$

$$\bar{v} = \sin(\varepsilon_6)u + \cos(\varepsilon_6)v. \tag{20}$$

**DISCUSSION**

Looking at the list of all point symmetries of the Fokker - Planck equation in two-dimensional Euclidean space, we see, that there is no simple way to get, for example, fundamental solution of PDE, using these symmetries. We have not at our disposal such an instrument, as scaling of independent variables $t$, $x$, $y$, $u$, $v$. The result (12-15) of action of symmetry group on trivial solution is not very interesting from physical point of view.

Indirect way of use of Galilean transformations (7) was demonstrated in [1]. The transformation was used for generalisation of solution, which was obtained in the form of exponent of quadratic form of space coordinates and velocities with time dependent coefficients.

There is need of further investigations of Fokker - Planck equation and its set of symmetries, which may lead to another physically interesting results. We can follow the scheme of [7] : to consider invariant solutions for some one-parameter group, thus reduce the independent variables number. To find for obtained in such a way equation all point symmetries - and so long.

In the work [7] this scheme was represented for equations of elasticity and plasticity.

**ACKNOWLEDGMENTS**

We wish to thank Jos A. M. Vermaseren from NIKHEF (the Dutch Institute for Nuclear and High-Energy Physics), for he made his symbolic computations program FORM release 3.1 available for download for non-commercial purposes (see [8]). This wonderful program makes difficult task of symmetries search more accessible.

---

## APPENDIX 1

The infinitesimal invariance criteria for PDE (1) is

$$\delta(\frac{\partial n}{\partial t}) + \delta u \frac{\partial n}{\partial x} + u\delta(\frac{\partial n}{\partial x}) + \delta v \frac{\partial n}{\partial y} + v\delta(\frac{\partial n}{\partial y}) - \quad\quad (A1\text{-}1)$$

$$-a\delta u \frac{\partial n}{\partial u} - au\delta(\frac{\partial n}{\partial u}) - a\delta v \frac{\partial n}{\partial v} - av\delta(\frac{\partial n}{\partial v}) - 2a\delta n - k(\delta(\frac{\partial^2 n}{\partial u^2}) + \delta(\frac{\partial^2 n}{\partial v^2})) = 0.$$

According to [2] (APPENDIX 1, eq. (A1-8) and (A1-18)), we have for variations of derivatives following expression:

$$\delta \frac{\partial n}{\partial x} = \frac{\partial}{\partial x}(\delta n) + \frac{\partial}{\partial n}(\delta n)\frac{\partial n}{\partial x} - \frac{\partial n}{\partial x}(\frac{\partial}{\partial x}(\delta x) + \frac{\partial}{\partial n}(\delta x)\frac{\partial n}{\partial x}) - \quad\quad (A1\text{-}2)$$

$$-\frac{\partial n}{\partial y}(\frac{\partial}{\partial x}(\delta y) + \frac{\partial}{\partial n}(\delta y)\frac{\partial n}{\partial x}) - \frac{\partial n}{\partial u}(\frac{\partial}{\partial x}(\delta u) + \frac{\partial}{\partial n}(\delta u)\frac{\partial n}{\partial x}) -$$

$$-\frac{\partial n}{\partial v}(\frac{\partial}{\partial x}(\delta v) + \frac{\partial}{\partial n}(\delta v)\frac{\partial n}{\partial x}) - \frac{\partial n}{\partial t}(\frac{\partial}{\partial x}(\delta t) + \frac{\partial}{\partial n}(\delta t)\frac{\partial n}{\partial x});$$

$$\delta \frac{\partial n}{\partial y} = \frac{\partial}{\partial y}(\delta n) + \frac{\partial}{\partial n}(\delta n)\frac{\partial n}{\partial y} - \frac{\partial n}{\partial x}(\frac{\partial}{\partial y}(\delta x) + \frac{\partial}{\partial n}(\delta x)\frac{\partial n}{\partial y}) - \quad\quad (A1\text{-}3)$$

$$-\frac{\partial n}{\partial y}(\frac{\partial}{\partial y}(\delta y) + \frac{\partial}{\partial n}(\delta y)\frac{\partial n}{\partial y}) - \frac{\partial n}{\partial u}(\frac{\partial}{\partial y}(\delta u) + \frac{\partial}{\partial n}(\delta u)\frac{\partial n}{\partial y}) -$$

$$-\frac{\partial n}{\partial v}(\frac{\partial}{\partial y}(\delta v) + \frac{\partial}{\partial n}(\delta v)\frac{\partial n}{\partial y}) - \frac{\partial n}{\partial t}(\frac{\partial}{\partial y}(\delta t) + \frac{\partial}{\partial n}(\delta t)\frac{\partial n}{\partial y});$$

$$\delta \frac{\partial n}{\partial u} = \frac{\partial}{\partial u}(\delta n) + \frac{\partial}{\partial n}(\delta n)\frac{\partial n}{\partial u} - \frac{\partial n}{\partial x}(\frac{\partial}{\partial u}(\delta x) + \frac{\partial}{\partial n}(\delta x)\frac{\partial n}{\partial u}) - \quad\quad (A1\text{-}4)$$

$$-\frac{\partial n}{\partial y}(\frac{\partial}{\partial u}(\delta y) + \frac{\partial}{\partial n}(\delta y)\frac{\partial n}{\partial u}) - \frac{\partial n}{\partial u}(\frac{\partial}{\partial u}(\delta u) + \frac{\partial}{\partial n}(\delta u)\frac{\partial n}{\partial u}) -$$

$$-\frac{\partial n}{\partial v}(\frac{\partial}{\partial u}(\delta v) + \frac{\partial}{\partial n}(\delta v)\frac{\partial n}{\partial u}) - \frac{\partial n}{\partial t}(\frac{\partial}{\partial u}(\delta t) + \frac{\partial}{\partial n}(\delta t)\frac{\partial n}{\partial u});$$

$$\delta \frac{\partial n}{\partial v} = \frac{\partial}{\partial v}(\delta n) + \frac{\partial}{\partial n}(\delta n)\frac{\partial n}{\partial v} - \frac{\partial n}{\partial x}(\frac{\partial}{\partial v}(\delta x) + \frac{\partial}{\partial n}(\delta x)\frac{\partial n}{\partial v}) - \quad\quad (A1\text{-}5)$$

$$-\frac{\partial n}{\partial y}(\frac{\partial}{\partial v}(\delta y) + \frac{\partial}{\partial n}(\delta y)\frac{\partial n}{\partial v}) - \frac{\partial n}{\partial u}(\frac{\partial}{\partial v}(\delta u) + \frac{\partial}{\partial n}(\delta u)\frac{\partial n}{\partial v}) -$$

$$-\frac{\partial n}{\partial v}(\frac{\partial}{\partial v}(\delta v) + \frac{\partial}{\partial n}(\delta v)\frac{\partial n}{\partial v}) - \frac{\partial n}{\partial t}(\frac{\partial}{\partial v}(\delta t) + \frac{\partial}{\partial n}(\delta t)\frac{\partial n}{\partial v});$$

$$\delta \frac{\partial n}{\partial t} = \frac{\partial}{\partial t}(\delta n) + \frac{\partial}{\partial n}(\delta n)\frac{\partial n}{\partial t} - \frac{\partial n}{\partial x}(\frac{\partial}{\partial t}(\delta x) + \frac{\partial}{\partial n}(\delta x)\frac{\partial n}{\partial t}) - \quad\quad (A1\text{-}6)$$

$$-\frac{\partial n}{\partial y}(\frac{\partial}{\partial t}(\delta y) + \frac{\partial}{\partial n}(\delta y)\frac{\partial n}{\partial t}) - \frac{\partial n}{\partial u}(\frac{\partial}{\partial t}(\delta u) + \frac{\partial}{\partial n}(\delta u)\frac{\partial n}{\partial t}) -$$

$$-\frac{\partial n}{\partial v}(\frac{\partial}{\partial t}(\delta v)+\frac{\partial}{\partial n}(\delta v)\frac{\partial n}{\partial t})-\frac{\partial n}{\partial t}(\frac{\partial}{\partial t}(\delta t)+\frac{\partial}{\partial n}(\delta t)\frac{\partial n}{\partial t});$$

$$\delta \frac{\partial^2 n}{\partial u^2} = \frac{\partial}{\partial n}(\delta n)\frac{\partial^2 n}{\partial u^2} - \frac{\partial^2 n}{\partial u \partial x}(\frac{\partial}{\partial u}(\delta x)+\frac{\partial}{\partial n}(\delta x)\frac{\partial n}{\partial u}) - \frac{\partial^2 n}{\partial u \partial y}(\frac{\partial}{\partial u}(\delta y)+\frac{\partial}{\partial n}(\delta y)\frac{\partial n}{\partial u}) - \quad (A1\text{-}7)$$

$$-\frac{\partial^2 n}{\partial u^2}(\frac{\partial}{\partial u}(\delta u)+\frac{\partial}{\partial n}(\delta u)\frac{\partial n}{\partial u}) - \frac{\partial^2 n}{\partial u \partial v}(\frac{\partial}{\partial u}(\delta v)+\frac{\partial}{\partial n}(\delta v)\frac{\partial n}{\partial u}) - \frac{\partial^2 n}{\partial t \partial u}(\frac{\partial}{\partial u}(\delta t)+\frac{\partial}{\partial n}(\delta t)\frac{\partial n}{\partial u}) -$$

$$-\frac{\partial^2 n}{\partial u^2}\frac{\partial n}{\partial x}\frac{\partial}{\partial n}(\delta x) - \frac{\partial^2 n}{\partial u^2}\frac{\partial n}{\partial y}\frac{\partial}{\partial n}(\delta y) - \frac{\partial^2 n}{\partial u^2}\frac{\partial n}{\partial u}\frac{\partial}{\partial n}(\delta u) - \frac{\partial^2 n}{\partial u^2}\frac{\partial n}{\partial v}\frac{\partial}{\partial n}(\delta v) - \frac{\partial^2 n}{\partial u^2}\frac{\partial n}{\partial t}\frac{\partial}{\partial n}(\delta t) +$$

$$+\frac{\partial^2}{\partial u^2}(\delta n) + \frac{\partial^2}{\partial n \partial u}(\delta n)\frac{\partial n}{\partial u} + \frac{\partial n}{\partial u}(\frac{\partial^2}{\partial n \partial u}(\delta n)+\frac{\partial^2}{\partial n^2}(\delta n)\frac{\partial n}{\partial u}) -$$

$$-\frac{\partial n}{\partial x}(\frac{\partial^2}{\partial u^2}(\delta x)+\frac{\partial^2}{\partial n \partial u}(\delta x)\frac{\partial n}{\partial u}) - \frac{\partial n}{\partial y}(\frac{\partial^2}{\partial u^2}(\delta y)+\frac{\partial^2}{\partial n \partial u}(\delta y)\frac{\partial n}{\partial u}) -$$

$$-\frac{\partial n}{\partial u}(\frac{\partial^2}{\partial u^2}(\delta u)+\frac{\partial^2}{\partial n \partial u}(\delta u)\frac{\partial n}{\partial u}) - \frac{\partial n}{\partial v}(\frac{\partial^2}{\partial u^2}(\delta v)+\frac{\partial^2}{\partial n \partial u}(\delta v)\frac{\partial n}{\partial u}) -$$

$$-\frac{\partial n}{\partial t}(\frac{\partial^2}{\partial u^2}(\delta t)+\frac{\partial^2}{\partial n \partial u}(\delta t)\frac{\partial n}{\partial u}) - \frac{\partial n}{\partial x}\frac{\partial n}{\partial u}(\frac{\partial^2}{\partial n \partial u}(\delta x)+\frac{\partial^2}{\partial n^2}(\delta x)\frac{\partial n}{\partial u}) -$$

$$-\frac{\partial n}{\partial y}\frac{\partial n}{\partial u}(\frac{\partial^2}{\partial n \partial u}(\delta y)+\frac{\partial^2}{\partial n^2}(\delta y)\frac{\partial n}{\partial u}) - \frac{\partial n}{\partial u}\frac{\partial n}{\partial u}(\frac{\partial^2}{\partial n \partial u}(\delta u)+\frac{\partial^2}{\partial n^2}(\delta u)\frac{\partial n}{\partial u}) -$$

$$-\frac{\partial n}{\partial v}\frac{\partial n}{\partial u}(\frac{\partial^2}{\partial n \partial u}(\delta v)+\frac{\partial^2}{\partial n^2}(\delta v)\frac{\partial n}{\partial u}) - \frac{\partial n}{\partial t}\frac{\partial n}{\partial u}(\frac{\partial^2}{\partial n \partial u}(\delta t)+\frac{\partial^2}{\partial n^2}(\delta t)\frac{\partial n}{\partial u}) -$$

$$-\frac{\partial^2 n}{\partial u \partial x}(\frac{\partial}{\partial u}(\delta x)+\frac{\partial}{\partial n}(\delta x)\frac{\partial n}{\partial u}) - \frac{\partial^2 n}{\partial u \partial y}(\frac{\partial}{\partial u}(\delta y)+\frac{\partial}{\partial n}(\delta y)\frac{\partial n}{\partial u}) -$$

$$-\frac{\partial^2 n}{\partial u^2}(\frac{\partial}{\partial u}(\delta u)+\frac{\partial}{\partial n}(\delta u)\frac{\partial n}{\partial u}) - \frac{\partial^2 n}{\partial u \partial v}(\frac{\partial}{\partial u}(\delta v)+\frac{\partial}{\partial n}(\delta v)\frac{\partial n}{\partial u}) - \frac{\partial^2 n}{\partial t \partial u}(\frac{\partial}{\partial u}(\delta t)+\frac{\partial}{\partial n}(\delta t)\frac{\partial n}{\partial u});$$

$$\delta \frac{\partial^2 n}{\partial v^2} = \frac{\partial}{\partial n}(\delta n)\frac{\partial^2 n}{\partial v^2} - \frac{\partial^2 n}{\partial v \partial x}(\frac{\partial}{\partial v}(\delta x)+\frac{\partial}{\partial n}(\delta x)\frac{\partial n}{\partial v}) - \frac{\partial^2 n}{\partial v \partial y}(\frac{\partial}{\partial v}(\delta y)+\frac{\partial}{\partial n}(\delta y)\frac{\partial n}{\partial v}) - \quad (A1\text{-}8)$$

$$-\frac{\partial^2 n}{\partial u \partial v}(\frac{\partial}{\partial v}(\delta u)+\frac{\partial}{\partial n}(\delta u)\frac{\partial n}{\partial v}) - \frac{\partial^2 n}{\partial v^2}(\frac{\partial}{\partial v}(\delta v)+\frac{\partial}{\partial n}(\delta v)\frac{\partial n}{\partial v}) - \frac{\partial^2 n}{\partial t \partial v}(\frac{\partial}{\partial v}(\delta t)+\frac{\partial}{\partial n}(\delta t)\frac{\partial n}{\partial v}) -$$

$$-\frac{\partial^2 n}{\partial v^2}\frac{\partial n}{\partial x}\frac{\partial}{\partial n}(\delta x) - \frac{\partial^2 n}{\partial v^2}\frac{\partial n}{\partial y}\frac{\partial}{\partial n}(\delta y) - \frac{\partial^2 n}{\partial v^2}\frac{\partial n}{\partial u}\frac{\partial}{\partial n}(\delta u) - \frac{\partial^2 n}{\partial v^2}\frac{\partial n}{\partial v}\frac{\partial}{\partial n}(\delta v) - \frac{\partial^2 n}{\partial v^2}\frac{\partial n}{\partial t}\frac{\partial}{\partial n}(\delta t) +$$

$$+\frac{\partial^2}{\partial v^2}(\delta n) + \frac{\partial^2}{\partial n \partial v}(\delta n)\frac{\partial n}{\partial v} + \frac{\partial n}{\partial v}(\frac{\partial^2}{\partial n \partial v}(\delta n)+\frac{\partial^2}{\partial n^2}(\delta n)\frac{\partial n}{\partial v}) -$$

$$-\frac{\partial n}{\partial x}(\frac{\partial^2}{\partial v^2}(\delta x)+\frac{\partial^2}{\partial n \partial v}(\delta x)\frac{\partial n}{\partial v}) - \frac{\partial n}{\partial y}(\frac{\partial^2}{\partial v^2}(\delta y)+\frac{\partial^2}{\partial n \partial v}(\delta y)\frac{\partial n}{\partial v}) -$$

$$-\frac{\partial n}{\partial u}(\frac{\partial^2}{\partial v^2}(\delta u)+\frac{\partial^2}{\partial n \partial v}(\delta u)\frac{\partial n}{\partial v}) - \frac{\partial n}{\partial v}(\frac{\partial^2}{\partial v^2}(\delta v)+\frac{\partial^2}{\partial n \partial v}(\delta v)\frac{\partial n}{\partial v}) -$$

$$-\frac{\partial n}{\partial t}(\frac{\partial^2}{\partial v^2}(\delta t)+\frac{\partial^2}{\partial n\partial v}(\delta t)\frac{\partial n}{\partial v})-\frac{\partial n}{\partial x}\frac{\partial n}{\partial v}(\frac{\partial^2}{\partial n\partial v}(\delta x)+\frac{\partial^2}{\partial n^2}(\delta x)\frac{\partial n}{\partial v})-$$

$$-\frac{\partial n}{\partial y}\frac{\partial n}{\partial v}(\frac{\partial^2}{\partial n\partial v}(\delta y)+\frac{\partial^2}{\partial n^2}(\delta y)\frac{\partial n}{\partial v})-\frac{\partial n}{\partial u}\frac{\partial n}{\partial v}(\frac{\partial^2}{\partial n\partial v}(\delta u)+\frac{\partial^2}{\partial n^2}(\delta u)\frac{\partial n}{\partial v})-$$

$$-\frac{\partial n}{\partial v}\frac{\partial n}{\partial v}(\frac{\partial^2}{\partial n\partial v}(\delta v)+\frac{\partial^2}{\partial n^2}(\delta v)\frac{\partial n}{\partial v})-\frac{\partial n}{\partial t}\frac{\partial n}{\partial v}(\frac{\partial^2}{\partial n\partial v}(\delta t)+\frac{\partial^2}{\partial n^2}(\delta t)\frac{\partial n}{\partial v})-$$

$$-\frac{\partial^2 n}{\partial v\partial x}(\frac{\partial}{\partial v}(\delta x)+\frac{\partial}{\partial n}(\delta x)\frac{\partial n}{\partial v})-\frac{\partial^2 n}{\partial v\partial y}(\frac{\partial}{\partial v}(\delta y)+\frac{\partial}{\partial n}(\delta y)\frac{\partial n}{\partial v})-$$

$$-\frac{\partial^2 n}{\partial u\partial v}(\frac{\partial}{\partial v}(\delta u)+\frac{\partial}{\partial n}(\delta u)\frac{\partial n}{\partial v})-\frac{\partial^2 n}{\partial v^2}(\frac{\partial}{\partial v}(\delta v)+\frac{\partial}{\partial n}(\delta v)\frac{\partial n}{\partial v})-\frac{\partial^2 n}{\partial t\partial v}(\frac{\partial}{\partial v}(\delta t)+\frac{\partial}{\partial n}(\delta t)\frac{\partial n}{\partial v});$$

We eliminate $\frac{\partial n}{\partial t}$ in (A1-1) using original equation

$$\frac{\partial n}{\partial t}=-\left(u\frac{\partial n}{\partial x}+v\frac{\partial n}{\partial y}-au\frac{\partial n}{\partial u}-av\frac{\partial n}{\partial v}-2an-k(\frac{\partial^2 n}{\partial u^2}+\frac{\partial^2 n}{\partial v^2})\right). \qquad (A1-9)$$

Collecting similar terms, we obtain following equations:

$\frac{\partial n}{\partial x}\frac{\partial n}{\partial u}$

$$-2ku\frac{\partial^2}{\partial n\partial u}(\delta t)+2k\frac{\partial^2}{\partial n\partial u}(\delta x)=0; \qquad (A1-10)$$

$\frac{\partial n}{\partial x}\frac{\partial n}{\partial u^2}$

$$-ku\frac{\partial^2}{\partial n^2}(\delta t)+k\frac{\partial^2}{\partial n^2}(\delta x)=0; \qquad (A1-11)$$

$\frac{\partial n}{\partial x}\frac{\partial n}{\partial v}$

$$-2ku\frac{\partial^2}{\partial n\partial v}(\delta t)+2k\frac{\partial^2}{\partial n\partial v}(\delta x)=0; \qquad (A1-12)$$

$\frac{\partial n}{\partial x}\frac{\partial n}{\partial v^2}$

$$-ku\frac{\partial^2}{\partial n^2}(\delta t)+k\frac{\partial^2}{\partial n^2}(\delta x)=0; \qquad (A1-13)$$

$\frac{\partial n}{\partial x}$

$$-auv\frac{\partial}{\partial v}(\delta t)+2aun\frac{\partial}{\partial n}(\delta t)+au\frac{\partial}{\partial u}(\delta x)-au^2\frac{\partial}{\partial u}(\delta t)+av\frac{\partial}{\partial v}(\delta x)- \qquad (A1-14)$$

$$-2an\frac{\partial}{\partial n}(\delta x)-ku\frac{\partial^2}{\partial u^2}(\delta t)-ku\frac{\partial^2}{\partial v^2}(\delta t)+k\frac{\partial^2}{\partial u^2}(\delta x)+k\frac{\partial^2}{\partial v^2}(\delta x)+$$

$$+uv\frac{\partial}{\partial y}(\delta t)-u\frac{\partial}{\partial x}(\delta x)+u\frac{\partial}{\partial t}(\delta t)+u^2\frac{\partial}{\partial x}(\delta t)-v\frac{\partial}{\partial y}(\delta x)+\delta u-\frac{\partial}{\partial t}(\delta x)=0;$$

$\dfrac{\partial n}{\partial y}\dfrac{\partial n}{\partial u}$

$$-2kv\dfrac{\partial^2}{\partial n\partial u}(\delta t)+2k\dfrac{\partial^2}{\partial n\partial u}(\delta y)=0; \tag{A1-15}$$

$\dfrac{\partial n}{\partial y}\dfrac{\partial n}{\partial u^2}$

$$-kv\dfrac{\partial^2}{\partial n^2}(\delta t)+k\dfrac{\partial^2}{\partial n^2}(\delta y)=0; \tag{A1-16}$$

$\dfrac{\partial n}{\partial y}\dfrac{\partial n}{\partial v}$

$$-2kv\dfrac{\partial^2}{\partial n\partial v}(\delta t)+2k\dfrac{\partial^2}{\partial n\partial v}(\delta y)=0; \tag{A1-17}$$

$\dfrac{\partial n}{\partial y}\dfrac{\partial n}{\partial v^2}$

$$-kv\dfrac{\partial^2}{\partial n^2}(\delta t)+k\dfrac{\partial^2}{\partial n^2}(\delta y)=0; \tag{A1-18}$$

$\dfrac{\partial n}{\partial y}$

$$-auv\dfrac{\partial}{\partial u}(\delta t)+au\dfrac{\partial}{\partial u}(\delta y)+2avn\dfrac{\partial}{\partial n}(\delta t)+av\dfrac{\partial}{\partial v}(\delta y)-av^2\dfrac{\partial}{\partial v}(\delta t)- \tag{A1-19}$$

$$-2an\dfrac{\partial}{\partial n}(\delta y)-kv\dfrac{\partial^2}{\partial u^2}(\delta t)-kv\dfrac{\partial^2}{\partial v^2}(\delta t)+k\dfrac{\partial^2}{\partial u^2}(\delta y)+k\dfrac{\partial^2}{\partial v^2}(\delta y)+$$

$$+uv\dfrac{\partial}{\partial x}(\delta t)-u\dfrac{\partial}{\partial x}(\delta y)-v\dfrac{\partial}{\partial y}(\delta y)v\dfrac{\partial}{\partial t}(\delta t)+v^2\dfrac{\partial}{\partial y}(\delta t)+\delta v-\dfrac{\partial}{\partial t}(\delta y)=0;$$

$\dfrac{\partial n}{\partial u}\dfrac{\partial n}{\partial v}$

$$2aku\dfrac{\partial^2}{\partial n\partial v}(\delta t)+2akv\dfrac{\partial^2}{\partial n\partial u}(\delta t)+2k\dfrac{\partial^2}{\partial n\partial u}(\delta v)+2k\dfrac{\partial^2}{\partial n\partial v}(\delta u)=0; \tag{A1-20}$$

$\dfrac{\partial n}{\partial u}\dfrac{\partial n}{\partial v^2}$

$$aku\dfrac{\partial^2}{\partial n^2}(\delta t)+k\dfrac{\partial^2}{\partial n^2}(\delta u)=0; \tag{A1-21}$$

$\dfrac{\partial n}{\partial u}\dfrac{\partial^2 n}{\partial u^2}$

$$2k\dfrac{\partial}{\partial n}(\delta u)+2k^2\dfrac{\partial^2}{\partial n\partial u}(\delta t)=0; \tag{A1-22}$$

$\dfrac{\partial n}{\partial u}\dfrac{\partial^2 n}{\partial v^2}$

$$2k^2\dfrac{\partial^2}{\partial n\partial u}(\delta t)=0; \tag{A1-23}$$

$\dfrac{\partial n}{\partial u}\dfrac{\partial^2 n}{\partial u\partial v}$

$$2k \frac{\partial}{\partial n} (\delta v) = 0; \qquad (A1\text{-}24)$$

$\frac{\partial n}{\partial u} \frac{\partial^2 n}{\partial u \partial x}$

$$2k \frac{\partial}{\partial n} (\delta x) = 0; \qquad (A1\text{-}25)$$

$\frac{\partial n}{\partial u} \frac{\partial^2 n}{\partial u \partial y}$

$$2k \frac{\partial}{\partial n} (\delta y) = 0; \qquad (A1\text{-}26)$$

$\frac{\partial n}{\partial u} \frac{\partial^2 n}{\partial t \partial u}$

$$2k \frac{\partial}{\partial n} (\delta t) = 0; \qquad (A1\text{-}27)$$

$\frac{\partial n}{\partial u}$

$$aku \frac{\partial^2}{\partial u^2} (\delta t) + aku \frac{\partial^2}{\partial v^2} (\delta t) + 4akn \frac{\partial^2}{\partial n \partial u} (\delta t) - auv \frac{\partial}{\partial y} (\delta t) + \qquad (A1\text{-}28)$$

$$+ au \frac{\partial}{\partial u} (\delta u) - au \frac{\partial}{\partial t} (\delta t) - au^2 \frac{\partial}{\partial x} (\delta t) + av \frac{\partial}{\partial v} (\delta u) - 2an \frac{\partial}{\partial n} (\delta u) - a\delta u + a^2 uv \frac{\partial}{\partial v} (\delta t) -$$

$$-2a^2 un \frac{\partial}{\partial n} (\delta t) + a^2 u^2 \frac{\partial}{\partial u} (\delta t) - 2k \frac{\partial^2}{\partial n \partial u} (\delta n) + k \frac{\partial^2}{\partial u^2} (\delta u) + k \frac{\partial^2}{\partial v^2} (\delta u) - u \frac{\partial}{\partial x} (\delta u) - v \frac{\partial}{\partial y} (\delta u) - \frac{\partial}{\partial t} (\delta u) = 0;$$

$\frac{\partial n}{\partial u^2} \frac{\partial n}{\partial v}$

$$akv \frac{\partial^2}{\partial n^2} (\delta t) + k \frac{\partial^2}{\partial n^2} (\delta v) = 0; \qquad (A1\text{-}29)$$

$\frac{\partial n}{\partial u^2} \frac{\partial^2 n}{\partial u^2}$

$$k^2 \frac{\partial^2}{\partial n^2} (\delta t) = 0; \qquad (A1\text{-}30)$$

$\frac{\partial n}{\partial u^2} \frac{\partial^2 n}{\partial v^2}$

$$k^2 \frac{\partial^2}{\partial n^2} (\delta t) = 0; \qquad (A1\text{-}31)$$

$\frac{\partial n}{\partial u^2}$

$$2aku \frac{\partial^2}{\partial n \partial u} (\delta t) + 2akn \frac{\partial^2}{\partial n^2} (\delta t) - k \frac{\partial^2}{\partial n^2} (\delta n) + 2k \frac{\partial^2}{\partial n \partial u} (\delta u) = 0; \qquad (A1\text{-}32)$$

$\frac{\partial n}{\partial u^3}$

$$aku \frac{\partial^2}{\partial n^2} (\delta t) + k \frac{\partial^2}{\partial n^2} (\delta u) = 0; \qquad (A1\text{-}33)$$

$\dfrac{\partial n}{\partial v}\dfrac{\partial^2 n}{\partial u^2}$

$$k^2 \dfrac{\partial^2}{\partial n \partial v}(\delta t) + k^2 \dfrac{\partial^2}{\partial n \partial v}(\delta t) = 0; \qquad (A1\text{-}34)$$

$\dfrac{\partial n}{\partial v}\dfrac{\partial^2 n}{\partial v^2}$

$$2k\dfrac{\partial}{\partial n}(\delta v) + 2k^2 \dfrac{\partial^2}{\partial n \partial v}(\delta t) = 0; \qquad (A1\text{-}35)$$

$\dfrac{\partial n}{\partial v}\dfrac{\partial^2 n}{\partial u \partial v}$

$$2k\dfrac{\partial}{\partial n}(\delta u) = 0; \qquad (A1\text{-}36)$$

$\dfrac{\partial n}{\partial v}\dfrac{\partial^2 n}{\partial v \partial x}$

$$2k\dfrac{\partial}{\partial n}(\delta x) = 0; \qquad (A1\text{-}37)$$

$\dfrac{\partial n}{\partial v}\dfrac{\partial^2 n}{\partial v \partial y}$

$$2k\dfrac{\partial}{\partial n}(\delta y) = 0; \qquad (A1\text{-}38)$$

$\dfrac{\partial n}{\partial v}\dfrac{\partial^2 n}{\partial t \partial v}$

$$2k\dfrac{\partial}{\partial n}(\delta t) = 0; \qquad (A1\text{-}39)$$

$\dfrac{\partial n}{\partial v}$

$$akv\dfrac{\partial^2}{\partial u^2}(\delta t) + akv\dfrac{\partial^2}{\partial v^2}(\delta t) + 4akn\dfrac{\partial^2}{\partial n \partial v}(\delta t) - \qquad (A1\text{-}40)$$

$$-auv\dfrac{\partial}{\partial x}(\delta t) + au\dfrac{\partial}{\partial u}(\delta v) + av\dfrac{\partial}{\partial v}(\delta v) - av\dfrac{\partial}{\partial t}(\delta t) -$$

$$-av^2\dfrac{\partial}{\partial y}(\delta t) - 2an\dfrac{\partial}{\partial n}(\delta v) - a\delta v + a^2 uv\dfrac{\partial}{\partial u}(\delta t) - 2a^2 vn\dfrac{\partial}{\partial n}(\delta t) +$$

$$+a^2 v^2 \dfrac{\partial}{\partial v}(\delta t) + k\dfrac{\partial^2}{\partial u^2}(\delta v) - 2k\dfrac{\partial^2}{\partial n \partial v}(\delta n) + k\dfrac{\partial^2}{\partial v^2}(\delta v) - u\dfrac{\partial}{\partial x}(\delta v) - v\dfrac{\partial}{\partial y}(\delta v) - \dfrac{\partial}{\partial t}(\delta v) = 0;$$

$\dfrac{\partial n}{\partial v^2}\dfrac{\partial^2 n}{\partial u^2}$

$$k^2 \dfrac{\partial^2}{\partial n^2}(\delta t) = 0; \qquad (A1\text{-}41)$$

$\dfrac{\partial n}{\partial v^2}\dfrac{\partial^2 n}{\partial v^2}$

$$k^2 \dfrac{\partial^2}{\partial n^2}(\delta t) = 0; \qquad (A1\text{-}42)$$

$\dfrac{\partial n}{\partial v^2}$

$$2akv \dfrac{\partial^2}{\partial n \partial v}(\delta t) + 2akn \dfrac{\partial^2}{\partial n^2}(\delta t) - k \dfrac{\partial^2}{\partial n^2}(\delta n) + 2k \dfrac{\partial^2}{\partial n \partial v}(\delta v) = 0; \qquad (A1\text{-}43)$$

$\dfrac{\partial n}{\partial v^3}$

$$akv \dfrac{\partial^2}{\partial n^2}(\delta t) + k \dfrac{\partial^2}{\partial n^2}(\delta v) = 0; \qquad (A1\text{-}44)$$

$\dfrac{\partial^2 n}{\partial u^2}$

$$aku \dfrac{\partial}{\partial u}(\delta t) + akv \dfrac{\partial}{\partial v}(\delta t) - 2akn \dfrac{\partial}{\partial n}(\delta t) - ku \dfrac{\partial}{\partial x}(\delta t) - \qquad (A1\text{-}45)$$

$$-kv \dfrac{\partial}{\partial y}(\delta t) + 2k \dfrac{\partial}{\partial u}(\delta u) - k \dfrac{\partial}{\partial t}(\delta t) + k^2 \dfrac{\partial^2}{\partial u^2}(\delta t) + k^2 \dfrac{\partial^2}{\partial v^2}(\delta t) = 0;$$

$\dfrac{\partial^2 n}{\partial v^2}$

$$aku \dfrac{\partial}{\partial u}(\delta t) + akv \dfrac{\partial}{\partial v}(\delta t) - 2akn \dfrac{\partial}{\partial n}(\delta t) - ku \dfrac{\partial}{\partial x}(\delta t) - \qquad (A1\text{-}46)$$

$$-kv \dfrac{\partial}{\partial y}(\delta t) + 2k \dfrac{\partial}{\partial v}(\delta v) - k \dfrac{\partial}{\partial t}(\delta t) + k^2 \dfrac{\partial^2}{\partial u^2}(\delta t) + k^2 \dfrac{\partial^2}{\partial v^2}(\delta t) = 0;$$

$\dfrac{\partial^2 n}{\partial u \partial v}$

$$2k \dfrac{\partial}{\partial v}(\delta u) + 2k \dfrac{\partial}{\partial u}(\delta v) = 0; \qquad (A1\text{-}47)$$

$\dfrac{\partial^2 n}{\partial u \partial x}$

$$2k \dfrac{\partial}{\partial u}(\delta x) = 0; \qquad (A1\text{-}48)$$

$\dfrac{\partial^2 n}{\partial u \partial y}$

$$2k \dfrac{\partial}{\partial u}(\delta y) = 0; \qquad (A1\text{-}49)$$

$\dfrac{\partial^2 n}{\partial v \partial x}$

$$2k \dfrac{\partial}{\partial v}(\delta x) = 0; \qquad (A1\text{-}50)$$

$\dfrac{\partial^2 n}{\partial v \partial y}$

$$2k \dfrac{\partial}{\partial v}(\delta y) = 0; \qquad (A1\text{-}51)$$

$\dfrac{\partial^2 n}{\partial t \partial u}$

$$2k \frac{\partial}{\partial u} (\delta t) = 0; \quad (A1\text{-}52)$$

$\frac{\partial^2 n}{\partial t \partial v}$

$$2k \frac{\partial}{\partial v} (\delta t) = 0; \quad (A1\text{-}53)$$

$$2akn \frac{\partial^2}{\partial u^2} (\delta t) + 2akn \frac{\partial^2}{\partial v^2} (\delta t) - 2aun \frac{\partial}{\partial x} (\delta t) - au \frac{\partial}{\partial u} (\delta n) - \quad (A1\text{-}54)$$

$$-2avn \frac{\partial}{\partial y} (\delta t) - av \frac{\partial}{\partial v} (\delta n) + 2an \frac{\partial}{\partial n} (\delta n) - 2an \frac{\partial}{\partial t} (\delta t) -$$

$$-2a\delta n + 2a^2 un \frac{\partial}{\partial u} (\delta t) + 2a^2 vn \frac{\partial}{\partial v} (\delta t) - 4a^2 n^2 \frac{\partial}{\partial n} (\delta t) - k \frac{\partial^2}{\partial u^2} (\delta n) -$$

$$-k \frac{\partial^2}{\partial v^2} (\delta n) + u \frac{\partial}{\partial x} (\delta n) + v \frac{\partial}{\partial y} (\delta n) + \frac{\partial}{\partial t} (\delta n) = 0.$$

From (A1-37 - A1-39), (A1-48 - A1-51) we see, that $\delta x = \delta x(x,y,t);\ \delta y = \delta y(x,y,t);$ $\delta t = \delta t(x,y,t)$. Using these expressions, we simplify the rest of equations (A1-10 - A1-54).

$$uv \frac{\partial}{\partial y} (\delta t) - u \frac{\partial}{\partial x} (\delta x) + u \frac{\partial}{\partial t} (\delta t) + u^2 \frac{\partial}{\partial x} (\delta t) - v \frac{\partial}{\partial y} (\delta x) + \delta u - \frac{\partial}{\partial t} (\delta x) = 0; \quad (A1\text{-}55)$$

$$uv \frac{\partial}{\partial x} (\delta t) - u \frac{\partial}{\partial x} (\delta y) - v \frac{\partial}{\partial y} (\delta y) + v \frac{\partial}{\partial t} (\delta t) + v^2 \frac{\partial}{\partial y} (\delta t) + \delta v - \frac{\partial}{\partial t} (\delta y) = 0; \quad (A1\text{-}56)$$

$$-auv \frac{\partial}{\partial y} (\delta t) + au \frac{\partial}{\partial u} (\delta u) - au \frac{\partial}{\partial t} (\delta t) - au^2 \frac{\partial}{\partial x} (\delta t) + \quad (A1\text{-}57)$$

$$+av \frac{\partial}{\partial v} (\delta u) - a\delta u - 2k \frac{\partial^2}{\partial n \partial u} (\delta n) + k \frac{\partial^2}{\partial u^2} (\delta u) +$$

$$+k \frac{\partial^2}{\partial v^2} (\delta u) - u \frac{\partial}{\partial x} (\delta u) - v \frac{\partial}{\partial y} (\delta u) - \frac{\partial}{\partial t} (\delta u) = 0;$$

$$-k \frac{\partial^2}{\partial n^2} (\delta n) = 0; \quad (A1\text{-}58)$$

$$-auv \frac{\partial}{\partial x} (\delta t) + au \frac{\partial}{\partial u} (\delta v) + av \frac{\partial}{\partial v} (\delta v) - av \frac{\partial}{\partial t} (\delta t) - \quad (A1\text{-}59)$$

$$-av^2 \frac{\partial}{\partial y} (\delta t) - a\delta v + k \frac{\partial^2}{\partial u^2} (\delta v) - 2k \frac{\partial^2}{\partial n \partial v} (\delta n) +$$

$$+k \frac{\partial^2}{\partial v^2} (\delta v) - u \frac{\partial}{\partial x} (\delta v) - v \frac{\partial}{\partial y} (\delta v) - \frac{\partial}{\partial t} (\delta v) = 0;$$

$$-k \frac{\partial^2}{\partial n^2} (\delta n) = 0; \quad (A1\text{-}60)$$

$$-ku\frac{\partial}{\partial x}(\delta t) - kv\frac{\partial}{\partial y}(\delta t) + 2k\frac{\partial}{\partial u}(\delta u) - k\frac{\partial}{\partial t}(\delta t) = 0; \tag{A1-61}$$

$$-ku\frac{\partial}{\partial x}(\delta t) - kv\frac{\partial}{\partial y}(\delta t) + 2k\frac{\partial}{\partial v}(\delta v) - k\frac{\partial}{\partial t}(\delta t) = 0; \tag{A1-62}$$

$$2k\frac{\partial}{\partial v}(\delta u) + 2k\frac{\partial}{\partial u}(\delta v) = 0; \tag{A1-63}$$

$$-2aun\frac{\partial}{\partial x}(\delta t) - au\frac{\partial}{\partial u}(\delta n) - 2avn\frac{\partial}{\partial y}(\delta t) - av\frac{\partial}{\partial v}(\delta n) + \tag{A1-64}$$

$$+2an\frac{\partial}{\partial n}(\delta n) - 2an\frac{\partial}{\partial t}(\delta t) - 2a\delta n - k\frac{\partial^2}{\partial u^2}(\delta n) -$$

$$-k\frac{\partial^2}{\partial v^2}(\delta n) + u\frac{\partial}{\partial x}(\delta n) + v\frac{\partial}{\partial y}(\delta n) + \frac{\partial}{\partial t}(\delta n) = 0.$$

From (A1-58) and (A1-60) we conclude, that

$$\delta n = A + nB; \tag{A1-65}$$

where $A = A(x, y, u, v, t), B = B(x, y, u, v, t)$. Using this expression, we simplify the rest of equations (A1-55 - A1-64).

$$uv\frac{\partial}{\partial y}(\delta t) - u\frac{\partial}{\partial x}(\delta x) + u\frac{\partial}{\partial t}(\delta t) + u^2\frac{\partial}{\partial x}(\delta t) - v\frac{\partial}{\partial y}(\delta x) + \delta u - \frac{\partial}{\partial t}(\delta x) = 0; \tag{A1-66}$$

$$uv\frac{\partial}{\partial x}(\delta t) - u\frac{\partial}{\partial x}(\delta y) - v\frac{\partial}{\partial y}(\delta y) + v\frac{\partial}{\partial t}(\delta t) + v^2\frac{\partial}{\partial y}(\delta t) + \delta v - \frac{\partial}{\partial t}(\delta y) = 0; \tag{A1-67}$$

$$-auv\frac{\partial}{\partial y}(\delta t) + au\frac{\partial}{\partial u}(\delta u) - au\frac{\partial}{\partial t}(\delta t) - au^2\frac{\partial}{\partial x}(\delta t) + \tag{A1-68}$$

$$+av\frac{\partial}{\partial v}(\delta u) - a\delta u + k\frac{\partial^2}{\partial u^2}(\delta u) + k\frac{\partial^2}{\partial v^2}(\delta u) - 2k\frac{\partial B}{\partial u} - u\frac{\partial}{\partial x}(\delta u) - v\frac{\partial}{\partial y}(\delta u) - \frac{\partial}{\partial t}(\delta u) = 0;$$

$$-auv\frac{\partial}{\partial x}(\delta t) + au\frac{\partial}{\partial u}(\delta v) + av\frac{\partial}{\partial v}(\delta v) - av\frac{\partial}{\partial t}(\delta t) - \tag{A1-69}$$

$$-av^2\frac{\partial}{\partial y}(\delta t) - a\delta v + k\frac{\partial^2}{\partial u^2}(\delta v) + k\frac{\partial^2}{\partial v^2}(\delta v) - 2k\frac{\partial B}{\partial v} - u\frac{\partial}{\partial x}(\delta v) - v\frac{\partial}{\partial y}(\delta v) - \frac{\partial}{\partial t}(\delta v) = 0;$$

$$-ku\frac{\partial}{\partial x}(\delta t) - kv\frac{\partial}{\partial y}(\delta t) + 2k\frac{\partial}{\partial u}(\delta u) - k\frac{\partial}{\partial t}(\delta t) = 0; \tag{A1-70}$$

$$-ku\frac{\partial}{\partial x}(\delta t) - kv\frac{\partial}{\partial y}(\delta t) + 2k\frac{\partial}{\partial v}(\delta v) - k\frac{\partial}{\partial t}(\delta t) = 0; \tag{A1-71}$$

$$2k\frac{\partial}{\partial v}(\delta u) + 2k\frac{\partial}{\partial u}(\delta v) = 0; \tag{A1-72}$$

$$-2au\frac{\partial}{\partial x}(\delta t) - au\frac{\partial B}{\partial u} - 2av\frac{\partial}{\partial y}(\delta t) - av\frac{\partial B}{\partial v} - 2a\frac{\partial}{\partial t}(\delta t) - \tag{A1-73}$$

$$-k\frac{\partial^2 B}{\partial u^2} - k\frac{\partial^2 B}{\partial v^2} + u\frac{\partial B}{\partial x} + v\frac{\partial B}{\partial y} + \frac{\partial B}{\partial t} = 0;$$

$$-au\frac{\partial A}{\partial u} - av\frac{\partial A}{\partial v} - 2aA - k\frac{\partial^2 A}{\partial u^2} - -k\frac{\partial^2 A}{\partial v^2} + u\frac{\partial A}{\partial x} + v\frac{\partial A}{\partial y} + \frac{\partial A}{\partial t} = 0. \qquad (A1\text{-}74)$$

Equation (A1-74) is simply Fokker - Planck equation for $A$.

We solve (A1-66 - A1-67) and find $\delta u, \delta v$

$$\delta u = -(uv\frac{\partial}{\partial y}(\delta t) - u\frac{\partial}{\partial x}(\delta x) + u\frac{\partial}{\partial t}(\delta t) + u^2\frac{\partial}{\partial x}(\delta t) - v\frac{\partial}{\partial y}(\delta x) - \frac{\partial}{\partial t}(\delta x)); \qquad (A1\text{-}76)$$

$$\delta v = -(uv\frac{\partial}{\partial x}(\delta t) - u\frac{\partial}{\partial x}(\delta y) - v\frac{\partial}{\partial y}(\delta y) + v\frac{\partial}{\partial t}(\delta t) + v^2\frac{\partial}{\partial y}(\delta t) - \frac{\partial}{\partial t}(\delta y)). \qquad (A1\text{-}77)$$

This gives for (A1-67 - A1-73)

$$-2auv\frac{\partial}{\partial y}(\delta t) - au\frac{\partial}{\partial t}(\delta t) - 2au^2\frac{\partial}{\partial x}(\delta t) - a\frac{\partial}{\partial t}(\delta x) - 2k\frac{\partial}{\partial x}(\delta t) - \qquad (A1\text{-}78)$$

$$-2k\frac{\partial B}{\partial u} + 2uv\frac{\partial^2}{\partial t\partial y}(\delta t) - 2uv\frac{\partial^2}{\partial x\partial y}(\delta x) + uv^2\frac{\partial^2}{\partial y^2}(\delta t) + u\frac{\partial^2}{\partial t^2}(\delta t) - 2u\frac{\partial^2}{\partial t\partial x}(\delta x) +$$

$$+2u^2 v\frac{\partial^2}{\partial x\partial y}(\delta t) + 2u^2\frac{\partial^2}{\partial t\partial x}(\delta t) - u^2\frac{\partial^2}{\partial x^2}(\delta x) + u^3\frac{\partial^2}{\partial x^2}(\delta t) - 2v\frac{\partial^2}{\partial t\partial y}(\delta x) - v^2\frac{\partial^2}{\partial y^2}(\delta x) - \frac{\partial^2}{\partial t^2}(\delta x) = 0;$$

$$-2auv\frac{\partial}{\partial x}(\delta t) - av\frac{\partial}{\partial t}(\delta t) - 2av^2\frac{\partial}{\partial y}(\delta t) - a\frac{\partial}{\partial t}(\delta y) - 2k\frac{\partial}{\partial y}(\delta t) - \qquad (A1\text{-}79)$$

$$-2k\frac{\partial B}{\partial v} + 2uv\frac{\partial^2}{\partial t\partial x}(\delta t) - 2uv\frac{\partial^2}{\partial x\partial y}(\delta y) + 2uv^2\frac{\partial^2}{\partial x\partial y}(\delta t) - 2u\frac{\partial^2}{\partial t\partial x}(\delta y) +$$

$$+u^2 v\frac{\partial^2}{\partial x^2}(\delta t) - u^2\frac{\partial^2}{\partial x^2}(\delta y) + v\frac{\partial^2}{\partial t^2}(\delta t) - 2v\frac{\partial^2}{\partial t\partial y}(\delta y) + 2v^2\frac{\partial^2}{\partial t\partial y}(\delta t) - v^2\frac{\partial^2}{\partial y^2}(\delta y) + v^3\frac{\partial^2}{\partial y^2}(\delta t) - \frac{\partial^2}{\partial t^2}(\delta y) = 0;$$

$$-5ku\frac{\partial}{\partial x}(\delta t) - 3kv\frac{\partial}{\partial y}(\delta t) + 2k\frac{\partial}{\partial x}(\delta x) - 3k\frac{\partial}{\partial t}(\delta t) = 0; \qquad (A1\text{-}80)$$

$$-3ku\frac{\partial}{\partial x}(\delta t) - 5kv\frac{\partial}{\partial y}(\delta t) + 2k\frac{\partial}{\partial y}(\delta y) - 3k\frac{\partial}{\partial t}(\delta t) = 0; \qquad (A1\text{-}81)$$

$$-2ku\frac{\partial}{\partial y}(\delta t) - 2kv\frac{\partial}{\partial x}(\delta t) + 2k\frac{\partial}{\partial y}(\delta x) + 2k\frac{\partial}{\partial x}(\delta y) = 0; \qquad (A1\text{-}82)$$

$$-2au\frac{\partial}{\partial x}(\delta t) - au\frac{\partial B}{\partial u} - 2av\frac{\partial}{\partial y}(\delta t) - av\frac{\partial B}{\partial v} - 2a\frac{\partial}{\partial t}(\delta t) - \qquad (A1\text{-}83)$$

$$-k\frac{\partial^2 B}{\partial u^2} - k\frac{\partial^2 B}{\partial v^2} + u\frac{\partial B}{\partial x} + v\frac{\partial B}{\partial y} + \frac{\partial B}{\partial t} = 0.$$

Now we can collect similar terms in (A1-80 - A1-82) and so split them into nine equations:

$$-5k\frac{\partial}{\partial x}(\delta t) = 0; \qquad (A1\text{-}84)$$

$$-3k\frac{\partial}{\partial y}(\delta t) = 0; \qquad (A1\text{-}85)$$

$$2k \frac{\partial}{\partial x} (\delta x) - 3k \frac{\partial}{\partial t} (\delta t) = 0; \qquad (A1\text{-}86)$$

$$-3k \frac{\partial}{\partial x} (\delta t) = 0; \qquad (A1\text{-}87)$$

$$-5k \frac{\partial}{\partial y} (\delta t) = 0; \qquad (A1\text{-}88)$$

$$2k \frac{\partial}{\partial y} (\delta y) - 3k \frac{\partial}{\partial t} (\delta t) = 0; \qquad (A1\text{-}89)$$

$$-2k \frac{\partial}{\partial y} (\delta t) = 0; \qquad (A1\text{-}90)$$

$$-2k \frac{\partial}{\partial x} (\delta t) = 0; \qquad (A1\text{-}91)$$

$$2k \frac{\partial}{\partial y} (\delta x) + 2k \frac{\partial}{\partial x} (\delta y) = 0. \qquad (A1\text{-}92)$$

From (A1-84 - A1-85), (A1-87 - A1-88), (A1-90 - A1-91) we see, that $\delta t = \delta t(t)$, which results in further simplifications

$$-au \frac{\partial}{\partial t}(\delta t) - a \frac{\partial}{\partial t}(\delta x) - 2k \frac{\partial B}{\partial u} - 2uv \frac{\partial^2}{\partial x \partial y}(\delta x) + u \frac{\partial^2}{\partial t^2}(\delta t) - 2u \frac{\partial^2}{\partial t \partial x}(\delta x) - \qquad (A1\text{-}93)$$

$$-u^2 \frac{\partial^2}{\partial x^2}(\delta x) - 2v \frac{\partial^2}{\partial t \partial y}(\delta x) - v^2 \frac{\partial^2}{\partial y^2}(\delta x) - \frac{\partial^2}{\partial t^2}(\delta x) = 0;$$

$$-av \frac{\partial}{\partial t}(\delta t) - a \frac{\partial}{\partial t}(\delta y) - 2k \frac{\partial B}{\partial v} - 2uv \frac{\partial^2}{\partial x \partial y}(\delta y) - 2u \frac{\partial^2}{\partial t \partial x}(\delta y) - \qquad (A1\text{-}94)$$

$$-u^2 \frac{\partial^2}{\partial x^2}(\delta y) + v \frac{\partial^2}{\partial t^2}(\delta t) - 2v \frac{\partial^2}{\partial t \partial y}(\delta y) - v^2 \frac{\partial^2}{\partial y^2}(\delta y) - \frac{\partial^2}{\partial t^2}(\delta y) = 0;$$

$$2k \frac{\partial}{\partial x}(\delta x) - 3k \frac{\partial}{\partial t}(\delta t) = 0; \qquad (A1\text{-}95)$$

$$2k \frac{\partial}{\partial y}(\delta y) - 3k \frac{\partial}{\partial t}(\delta t) = 0; \qquad (A1\text{-}96)$$

$$2k \frac{\partial}{\partial y}(\delta x) + 2k \frac{\partial}{\partial x}(\delta y) = 0; \qquad (A1\text{-}97)$$

$$-au \frac{\partial B}{\partial u} - av \frac{\partial B}{\partial v} - 2a \frac{\partial}{\partial t}(\delta t) - k \frac{\partial^2 B}{\partial u^2} - k \frac{\partial^2 B}{\partial v^2} + u \frac{\partial B}{\partial x} + v \frac{\partial B}{\partial y} + \frac{\partial B}{\partial t} = 0. \qquad (A1\text{-}98)$$

We integrate (A1-95 - A1-96) and find

$$\delta x = C + 3/2 x \frac{\partial}{\partial t}(\delta t); \qquad (A1\text{-}99)$$

$$\delta y = D + 3/2 y \frac{\partial}{\partial t}(\delta t); \qquad (A1\text{-}100)$$

where $C = C(y, t), D = D(x, t)$. We substitute these expressions to (A1-93 - A1-94), (A1-97 - A1-98) and obtain

$$-3/2ax \frac{\partial^2}{\partial t^2}(\delta t) - au \frac{\partial}{\partial t}(\delta t) - a \frac{\partial C}{\partial t} - 2k \frac{\partial B}{\partial u} - 3/2x \frac{\partial^3}{\partial t^3}(\delta t) - \quad (A1\text{-}101)$$

$$-2u \frac{\partial^2}{\partial t^2}(\delta t) - 2v \frac{\partial^2 C}{\partial t \partial y} - v^2 \frac{\partial^2 C}{\partial y^2} - \frac{\partial^2 C}{\partial t^2} = 0;$$

$$-3/2ay \frac{\partial^2}{\partial t^2}(\delta t) - av \frac{\partial}{\partial t}(\delta t) - a \frac{\partial D}{\partial t} - 2k \frac{\partial B}{\partial v} - 3/2y \frac{\partial^3}{\partial t^3}(\delta t) - \quad (A1\text{-}102)$$

$$-2u \frac{\partial^2 D}{\partial t \partial x} - u^2 \frac{\partial^2 D}{\partial x^2} - 2v \frac{\partial^2}{\partial t^2}(\delta t) - \frac{\partial^2 D}{\partial t^2} = 0;$$

$$2k \frac{\partial C}{\partial y} + 2k \frac{\partial D}{\partial x} = 0; \quad (A1\text{-}103)$$

$$-au \frac{\partial B}{\partial u} - av \frac{\partial B}{\partial v} - 2a \frac{\partial}{\partial t}(\delta t) - k \frac{\partial^2 B}{\partial u^2} - k \frac{\partial^2 B}{\partial v^2} + u \frac{\partial B}{\partial x} + v \frac{\partial B}{\partial y} + \frac{\partial B}{\partial t} = 0. \quad (A1\text{-}104)$$

We find $\frac{\partial B}{\partial u}$ from (A1-101) and $\frac{\partial B}{\partial v}$ from (A1-102):

$$\frac{\partial B}{\partial u} = \frac{1}{k}(-3/4ax \frac{\partial^2}{\partial t^2}(\delta t) - 1/2au \frac{\partial}{\partial t}(\delta t) - 1/2a \frac{\partial C}{\partial t} - 3/4x \frac{\partial^3}{\partial t^3}(\delta t) - \quad (A1\text{-}105)$$

$$-u \frac{\partial^2}{\partial t^2}(\delta t) - v \frac{\partial^2 C}{\partial t \partial y} - 1/2v^2 \frac{\partial^2 C}{\partial y^2} - 1/2 \frac{\partial^2 C}{\partial t^2});$$

$$\frac{\partial B}{\partial v} = \frac{1}{k}(-3/4ay \frac{\partial^2}{\partial t^2}(\delta t) - 1/2av \frac{\partial}{\partial t}(\delta t) - 1/2a \frac{\partial D}{\partial t} - 3/4y \frac{\partial^3}{\partial t^3}(\delta t) - \quad (A1\text{-}106)$$

$$-u \frac{\partial^2 D}{\partial t \partial x} - 1/2u^2 \frac{\partial^2 D}{\partial x^2} - v \frac{\partial^2}{\partial t^2}(\delta t) - 1/2 \frac{\partial^2 D}{\partial t^2}).$$

Differentiating (A1-105) by $v$ we have

$$\frac{\partial^2 B}{\partial u \partial v} = \frac{1}{k}(-v \frac{\partial^2 C}{\partial y^2} - \frac{\partial^2 C}{\partial t \partial y}); \quad (A1\text{-}107)$$

Differentiating (A1-106) by $u$ we have

$$\frac{\partial^2 B}{\partial u \partial v} = \frac{1}{k}(-u \frac{\partial^2 D}{\partial x^2} - \frac{\partial^2 D}{\partial t \partial x}). \quad (A1\text{-}108)$$

We know, that $C = C(y, t), D = D(x, t)$ and so we conclude from (A1-107 - A1-108)

$$C = C_1 y + E; \quad (A1\text{-}109)$$

$$D = C_2 x + F; \quad (A1\text{-}110)$$

where $E = E(t), F = F(t), C_1 + C_2 = 0$.

We find derivatives of $B$.

$$\frac{\partial B}{\partial u} = \frac{1}{2k}(3/2ax \frac{\partial^2}{\partial t^2}(\delta t) + au \frac{\partial}{\partial t}(\delta t) + a \frac{\partial E}{\partial t} - 3/2x \frac{\partial^3}{\partial t^3}(\delta t) - 2u \frac{\partial^2}{\partial t^2}(\delta t) - \frac{\partial^2 E}{\partial t^2}); \quad (A1\text{-}111)$$

$$\frac{\partial B}{\partial v} = \frac{1}{2k} (3/2ay \frac{\partial^2}{\partial t^2} (\delta t) + av \frac{\partial}{\partial t} (\delta t) + a \frac{\partial F}{\partial t} - 3/2y \frac{\partial^3}{\partial t^3} (\delta t) - 2v \frac{\partial^2}{\partial t^2} (\delta t) - \frac{\partial^2 F}{\partial t^2}); \quad (A1\text{-}112)$$

$$\frac{\partial^2 B}{\partial u \partial v} = \frac{\partial^2 B}{\partial v \partial u} = 0. \quad (A1\text{-}113)$$

Integration of (A1-111 - A1-112) gives

$$B = G + \frac{1}{2k} (3/2uax \frac{\partial^2}{\partial t^2} (\delta t) + 1/2au^2 \frac{\partial}{\partial t} (\delta t) + ua \frac{\partial E}{\partial t} - 3/2ux \frac{\partial^3}{\partial t^3} (\delta t) - u^2 \frac{\partial^2}{\partial t^2} (\delta t) - u \frac{\partial^2 E}{\partial t^2}) + \quad (A1\text{-}114)$$

$$+ \frac{1}{2k} (3/2vay \frac{\partial^2}{\partial t^2} (\delta t) + 1/2av^2 \frac{\partial}{\partial t} (\delta t) + va \frac{\partial F}{\partial t} - 3/2vy \frac{\partial^3}{\partial t^3} (\delta t) - v^2 \frac{\partial^2}{\partial t^2} (\delta t) - v \frac{\partial^2 F}{\partial t^2});$$

where $G = G(x, y, t)$.

Substitution of (A1-114) to (A1-98), collecting and equating to zero similar terms by $u, v$ gives

$$3/4a^2 k^{-1} x \frac{\partial^2}{\partial t^2} (\delta t) + 1/2a^2 k^{-1} \frac{\partial E}{\partial t} - 3/4k^{-1} x \frac{\partial^4}{\partial t^4} (\delta t) - 1/2k^{-1} \frac{\partial^3 E}{\partial t^3} + \frac{\partial G}{\partial x} = 0; \quad (A1\text{-}115)$$

$$1/2a^2 k^{-1} \frac{\partial}{\partial t} (\delta t) - 5/4k^{-1} \frac{\partial^3}{\partial t^3} (\delta t) = 0; \quad (A1\text{-}116)$$

$$3/4a^2 k^{-1} y \frac{\partial^2}{\partial t^2} (\delta t) + 1/2a^2 k^{-1} \frac{\partial F}{\partial t} - 3/4k^{-1} y \frac{\partial^4}{\partial t^4} (\delta t) - 1/2k^{-1} \frac{\partial^3 F}{\partial t^3} + \frac{\partial G}{\partial y} = 0; \quad (A1\text{-}117)$$

$$1/2a^2 k^{-1} \frac{\partial}{\partial t} (\delta t) - 5/4k^{-1} \frac{\partial^3}{\partial t^3} (\delta t) = 0; \quad (A1\text{-}118)$$

$$-a \frac{\partial}{\partial t} (\delta t) + 2 \frac{\partial^2}{\partial t^2} (\delta t) + \frac{\partial G}{\partial t} = 0; \quad (A1\text{-}119)$$

Integrate (A1-115), (A1-117) and obtain following expression ($H = H(t)$):

$$G = H - \frac{1}{k} \left( 3/8a^2 x^2 \frac{\partial^2}{\partial t^2} (\delta t) + 1/2xa^2 \frac{\partial E}{\partial t} - 3/8x^2 \frac{\partial^4}{\partial t^4} (\delta t) - 1/2x \frac{\partial^3 E}{\partial t^3} \right) - \quad (A1\text{-}120)$$

$$- \frac{1}{k} \left( 3/8a^2 y^2 \frac{\partial^2}{\partial t^2} (\delta t) + 1/2ya^2 \frac{\partial F}{\partial t} - 3/8y^2 \frac{\partial^4}{\partial t^4} (\delta t) - 1/2y \frac{\partial^3 F}{\partial t^3} \right).$$

Substitution of (A1-120) to (A1-119), collecting and equating to zero terms by $x, y$ gives

$$-1/2a^2 k^{-1} \frac{\partial^2 E}{\partial t^2} + 1/2k^{-1} \frac{\partial^4 E}{\partial t^4} = 0; \quad (A1\text{-}121)$$

$$-3/8a^2 k^{-1} \frac{\partial^3}{\partial t^3} (\delta t) + 3/8k^{-1} \frac{\partial^5}{\partial t^5} (\delta t) = 0; \quad (A1\text{-}122)$$

$$-1/2a^2 k^{-1} \frac{\partial^2 F}{\partial t^2} + 1/2k^{-1} \frac{\partial^4 F}{\partial t^4} = 0; \quad (A1\text{-}123)$$

$$-3/8a^2 k^{-1} \frac{\partial^3}{\partial t^3} (\delta t) + 3/8k^{-1} \frac{\partial^5}{\partial t^5} (\delta t) = 0; \quad (A1\text{-}124)$$

$$-a \frac{\partial}{\partial t} (\delta t) + 2 \frac{\partial^2}{\partial t^2} (\delta t) + \frac{\partial H}{\partial t} = 0. \quad (A1\text{-}125)$$

From (A1-116), (A1-118), (A1-122), (A1-124) we conclude, that

$$\delta t = const = C_3. \tag{A1-126}$$

From (A1-121), (A1-123) we conclude, that

$$E = C_4 + C_5 t + C_6 e^{-at} + C_7 e^{at}; \tag{A1-127}$$

$$F = C_8 + C_9 t + C_{10} e^{-at} + C_{11} e^{at}; \tag{A1-128}$$

From (A1-125) and (A1-126) we see, that

$$H = C_{12}. \tag{A1-129}$$

We obtain using (A1-126 - A1-129) and backward substitution final expressions for variations:

$$\delta n = -1/2 a k^{-1} u n C_5 - 1/2 a k^{-1} v n C_9 - 1/2 a^2 k^{-1} x n C_5 - \tag{A1-130}$$

$$-1/2 a^2 k^{-1} y n C_9 - a^2 k^{-1} u n C_7 e^{at} - a^2 k^{-1} v n C_{11} e^{at} + n C_{12} + A;$$

$$\delta x = y C_1 + t C_5 + C_4 + C_6 e^{-at} + C_7 e^{at}; \tag{A1-131}$$

$$\delta y = x C_2 + t C_9 + C_8 + C_{10} e^{-at} + C_{11} e^{at}; \tag{A1-132}$$

$$\delta u = -a C_6 e^{-at} + a C_7 e^{at} + v C_1 + C_5; \tag{A1-133}$$

$$\delta v = -a C_{10} e^{-at} + a C_{11} e^{at} + u C_2 + C_9; \tag{A1-134}$$

$$\delta t = C_3. \tag{A1-135}$$

This ends calculations.